\documentclass[aps,a4,twocolumn,prd,preprint,10pt,superscriptaddress]{revtex4}

\usepackage{graphicx}

\begin{document}

\title{Statistical linearizations for stochastically quantized fields}

\date{\today}

\author{Maciej Janowicz}
\email{mjanow@ifpan.edu.pl}
\affiliation{Institute of Physics, 
Carl von Ossietzky Universit\"at, D - 26111 Oldenburg, Germany}
\affiliation{Faculty of Computer Science, Warsaw University of Life Sciences - SGGW,
ul. Nowoursynowska 159, 02-786 Warsaw, Poland}

\author{Arkadiusz Or{\l}owski}
\affiliation{Institute of Physics of the Polish Academy of Sciences,
Aleja Lotnik\'ow 32/46, 02-668 Warsaw, Poland}
\affiliation{Faculty of Computer Science, Warsaw University of Life Sciences - SGGW,
ul. Nowoursynowska 159, 02-786 Warsaw, Poland}

\begin{abstract}
The statistical linearization method known in nonlinear mechanics and random
vibrations theory has been applied to stochastically quantized fields
in finite temperature. It has been shown that even in its simplest form the method 
yields convenient implicit equations for the self-energy, equivalent to the 
Dyson-Schwinger equations resulting from the summation of infinite number
of perturbative diagrams. Three examples have been provided:
the quantum anharmonic oscillator, the scalar $\phi^{4}$ theory in three spatial dimension, 
and the Bose-Hubbard model. The Ramanujan summation has been used to deal with divergent 
integrals and series.
\end{abstract}

\keywords{stochastic quantization; quantum scalar field; Bose-Hubbard model; statistical linearization}

\pacs{11.10.Wx; 11.15.Tk; 64.60.De; 02.50.Ey}

\maketitle

% \linenumbers

\section{Introduction}
\label{Intro}

The method of stochastic quantization invented in the end of seventies 
\cite{PW} has immediately grasped interest of several members of the field-theoretical
community, please see \cite{DH,Namiki,MS} for reviews. The method has offered a version of the quantum theory 
in which one has to deal with $c$-number fields (instead of operators or path integrals) 
at the expense of introduction of another temporal dimension and Langevin forces.
A very considerable amount of work has been applied to develop the theory and 
run simulations. In fact, it is the possibility of efficient simulations that has
strongly stimulated the research in this method. Initially, it has been worked
out as a way to obtain Euclidean vacuum expectation values of the (products of) quantum
fields. Later, it has been demonstrated that the method can also work in the Minkowski
space-time \cite{HR,Gozzi}. What is more, it has been shown that the stochastic
quantization can also be used to obtain transition amplitudes between quantum
states in simple quantum mechanical systems \cite{HN,YN}.

In spite of the initially vigorous development, the interest in stochastic quantization
diminished considerably in the nineties. The present authors believe that this happened
because of the following reasons: (i) the method has been worked out with the hope
to offer efficient simulations of Yang-Mills fields, especially in quantum
chromodynamics (QCD). However, the computational QCD has had in its disposal 
very efficient unrelated methods, and stochastic quantization has not seemed to offer
much more advantage from the computational point of view; (ii) there have been
serious difficulties in the simulations in Minkowski (rather than Euclidean) space
due to the instabilities caused by the Langevin forces; (iii) it appears that the 
non-perturbative analytic or semi-analytic methods have been somewhat underdeveloped
in the stochastic-quantization context - the only works on the subject of which are aware
are the variational approaches of \cite{AD} and \cite{BGG,BG1} and Hartree-like 
technique of \cite{BG2} (the above four papers are virtually forgotten); 
(iv) the method has not been sufficiently tested on simpler quantum systems - 
for instance, one would expect that before difficult lattice QCD simulations are launched, 
the stochastic quantization should first demonstrate that it allows to get, say, 
correct values of the ground-state energy of the helium atom.

The objective of the paper is to address the point (iii) and demonstrate that the stochastic quantization 
can very easily import well-known non-perturbative techniques from the non-linear mechanics
as well as the physics of random vibrations. Here, we use the simplest of those methods,
namely, that of statistical linearization \cite{Booton,Kazakov,Caughey,AU,Ito}, please see also
\cite{RS,Socha} for reviews. It allows one to replace a non-linear term in a Langevin equation 
with such a linear term that the expectation value of the resulting error is minimized. 
However, there exist also several other linearization criteria. We mention here the minimization
of the energy error, equality of mean-square energies, and equality of mean-square
system functions (i.e. the deterministic parts of the right-hand sides of corresponding Langevin equations).

Our analyses provided below may be classified as a specfic version of Hartree-like approach.
The equations of motion in the additional time-like variables are linearized
and the linear parameters which enter them are found variationally on assuming a Gaussian distribution 
of fields. Since the Hartree approaches have been discussed in multitude of papers,
there is a natural question whether it is profitable to investigate it again.
The present authors believe that it is indeed the case because the simplest
version of statistical linearization presented here is only a necessary first step
towards developing more sophisticated and non-trivial approximation schemes based 
on the so-called higher-order linearization as well as building statistically equivalent non-linear
solvable models. Less importantly, we have not found
any discussion of a Hartree-like approximation in the context of stochastic quantization
except of \cite{BG2} where it has been introduced, so to say, ``by hand" rather than with
the help of a systematic procedure like the one described below.

The main part of the paper is organized as follows. To introduce the technique of statistical
linearization in the stochastic-quantization context, the quantum anharmonic oscillator
in non-zero temperature is considered in Section 2. Temperature-dependent corrections to its
frequency are obtained. In Section 3, we consider the Langevin equations for $\phi^{4}$ theory
in three spatial dimensions and finite temperature. We show that the statistical linearization 
technique provides us with the correct self-consistent equation for the self-energy. 
Its zeroth-temperature limit is analyzed using the Ramanujan summation of divergent series. 
Section 4 contains an analysis of the Bose-Hubbard model from the point of view of stochastic quantization. 
The Langevin equations are again linearized and Dyson-Schwinger-like equation for the self-energy is
obtained. Section 5 contains some concluding remarks.

\section{Anharmonic oscillator}
\label{Dwa}
 
Let us consider the anharmonic oscillator described by the following Euclidean action (confer \cite{NK}):

\begin{equation}
S_{E} = \int_{0}^{\beta {\hbar}} \left[ \frac{1}{2} m {\dot x}^{2} + \frac{1}{2} m \omega^{2} x^2
+ \frac{m \lambda}{4!} x^{4}\right] d \tau,
\end{equation}
Ito
where $\tau$ is the Euclidean time, $x$ is the position of the oscillator, $m$ - its mass, $\omega$
denotes the frequency of the free oscillations, $\beta$ is inverse temperature, 
and ${\lambda}$ parametrizes the strength of nonlinearity.

According to the prescription of the stochastic quantization, $x$ should satisfy the equation:

$$
\frac{d x}{d s} = -\frac{\delta S_{E}}{\delta x} + \eta(s, \tau),
$$

where $\delta / \delta x$ denotes the functional derivative, and $s$ is an additional independent
variable of the temporal character while $\eta(s, \tau)$ is a Gaussian Langevin ``force" which satisfies 
the following conditions:

\begin{equation}
\langle \eta(s, t) \rangle = 0, \;\;\;\;\; 
\langle \eta(s, \tau) \eta(s^{\prime}, \tau^{\prime}) \rangle = 
2 {\hbar} \delta(s - s^{\prime}) \delta(t - t^{\prime}).
\end{equation}

Explicitly, the pseudo-dynamics in ``time" $s$ is governed by the following non-linear 
stochastic diffusion equation:

\begin{equation}
\label{AOmain}
\frac{d x}{d s} = m \frac{d^{2} x}{d \tau^{2}} - m \omega^{2} x - m \frac{\lambda}{3!} x^3
+ \eta(s, \tau).
\end{equation}

It has been proved \cite{PW,DH} that the Euclidean, ground-state correlation functions 
$\langle x(t_{1}) x(t_{2})... x(t_{n}) \rangle$ can be 
obtained by taking the limit:

$$
lim_{s \rightarrow \infty} \langle x(t_{1}) x(t_{2})... x(t_{n}) \rangle _{\eta},
$$

where the subscript $\eta$ denotes the averaging over the stochastic forces $\eta$.
Taking the limit is equivalent to the averaging by using a stationary solution
to the Fokker-Planck equation corresponding to the Langevin equation (\ref{AOmain}).

We look for the stationary ($s \rightarrow \infty$) solution of Eq.(\ref{AOmain}) subject to periodic boundary
condition in the Euclidean time $\tau$ to include finite temperature:

$$
x(s, 0) = x(s, \beta {\hbar}).
$$  

By ``stationary" solution" of (\ref{AOmain}) we mean the inhomogeneous part of the general
solution corresponding to $s$ sufficiently large that the initial values of $x(s)$ give 
vanishing contribution as they are exponentially damped.

Due to the boundary conditions, it is reasonable to expand:

$$
x(s, \tau) = \sum_{\nu = -\infty}^{\infty} x_{\nu}(s) u_{\nu}(\tau), 
$$

where 

$$
u_{\nu}(\tau) = \frac{1}{\sqrt{\beta {\hbar}}} \exp(2 \pi i \nu \tau/\beta {\hbar}) = 
\frac{1}{\sqrt{\beta {\hbar}}} \exp(i \omega_{\nu} \tau).
$$

The stochastic force can be expanded in the same way:

$$
\eta(s, \tau) = \sum_{\nu = -\infty}^{\infty} \eta_{\nu}(s) u_{\nu}(\tau), 
$$

with $\langle \eta_{\nu}(s) \eta_{\rho}(s^{\prime}) \rangle = 
(2 {\hbar}) \delta_{\nu, -\rho} \delta(s - s^{\prime})$.

The amplitudes $x_{\nu}(s)$ satisfy:

\begin{equation}
\label{AO2}
\frac{d x_{\nu}}{d s} = -m (\omega_{\nu}^2 + \omega^{2}) x_{\nu} - m \frac{{\lambda}}{3! \beta {\hbar}}
\sum_{\rho, \sigma} x_{\rho} x_{\sigma} x_{\nu - \rho -\sigma} + \eta_{\nu}(s).
\end{equation} 

Now, we are in position to look for an equivalent linear system. It has to take the form:

\begin{equation}
\label{AO3}
\frac{d x_{\nu}}{d s} + m \Omega_{\nu}^{2} x_{\nu} = \eta_{\nu}(s).
\end{equation}

The statistical linearization technique in its simplest version consists of subtracting the equations 
(\ref{AO2} - \ref{AO3}), taking the square of the resulting error, summing over all $\nu$ and taking
the expectation value. This expectation value should be computed with respect to the distribution
function of $x_{\nu}$ obtained by solving the full nonlinear system. As it is, however, not available,
we approximate that expectation value by its value obtained from the linearized system (\ref{AO3}).
Thus, we need to minimize the expression:

\begin{equation}
E = m \sum_{\nu} \langle | (\omega_{\nu}^2 + \omega^{2}) x_{\nu} + \frac{{\lambda}}{3! \beta {\hbar}} 
\sum_{\rho, \sigma} x_{\rho} x_{\sigma} x_{\nu - \rho - \sigma} - \Omega_{\nu}^{2} x_{\nu} |^{2}
\rangle
\end{equation}

The absolute value in the above expression appears naturally because $x_{\nu}$ are, naturally, complex;
they must satisfy, however, the condition $x_{-\nu} = x_{\nu}^{\star}$ because $x$ itself is real. 
Differentiating with respect to $\Omega_{\nu}$ and equating to zero the obtained derivative 
$\partial E/ \partial \Omega_{\mu}^{2}$ gives the following implicit equation for $\Omega_{\mu}^{2}$:

\begin{equation}
\Omega_{\mu}^{2} = \omega_{\mu}^{2} + \omega^{2} + \frac{{\lambda}}{3! \beta {\hbar}}
\frac{\sum_{\rho, \sigma} \langle x_{-\mu} x_{\rho} x_{\sigma} x_{\mu - \rho - \sigma}
\rangle}{\langle x_{\mu} x_{-\mu} \rangle}
\end{equation}

For Gaussian distribution of $x_{\mu}$, the above ratio of moments can be simplified \cite{RS}:

\begin{equation}
\Omega_{\mu}^{2} = \omega_{\mu}^{2} + \omega^{2} + \frac{{\lambda}}{3! \beta {\hbar}}
\langle \frac{\partial}{\partial x_{\mu}} \sum_{\rho, \sigma} 
x_{\rho} x_{\sigma} x_{\mu - \rho - \sigma}
\rangle,
\end{equation}

so that $\Omega_{\mu}^{2}$ depends only on the second moments of $x_{\mu}$:

\begin{equation}
\Omega_{\mu}^{2} = \omega_{\mu}^{2} + \omega^{2} + \frac{\lambda}{2 \beta {\hbar}} 
\sum_{\nu} \langle x_{\nu} x_{-\nu} \rangle. 
\end{equation}

Now, the system (\ref{AO3}) can be solved immediately to give for $s \rightarrow \infty$:

$$
\langle x_{\nu} x_{-\nu} \rangle = \frac{\hbar}{m \Omega_{\nu}^{2}}
$$

Let us now write $\Omega_{\nu}^{2} = \omega_{\nu}^{2} + \omega^{2} + \Pi$. Then the ``self-energy"
$\Pi$ satisfies the self-consistent equation:

\begin{equation}
\Pi = \frac{\lambda}{2 m \beta} \sum_{\nu} \frac{1}{\omega_{\nu}^{2} + \omega^{2} + \Pi}
\end{equation}

Performing now the frequency summation we find that $\Pi$ must satisfy:

\begin{equation}
\Pi = \frac{{\hbar} \lambda}{2 m \sqrt{\omega^{2} + \Pi}} \left( 1 + 2 n_{B}(\sqrt{\omega^{2} + \Pi}) \right),
\end{equation}

where $n_{B}(x)$ is the Bose-Einstein factor $(\exp(\beta {\hbar} x) - 1)^{-1}$.

In Fig. 1 there is a shaded contour plot of the dependence of the $\Pi/\omega^{2}$
on dimensionless coupling constant $L = {\hbar} \lambda/(2 m \omega^{3})$ and dimensionless
inverse temperature $b = \beta {\hbar} \omega$.

\begin{figure}
\begin{center}
\includegraphics[width = 10cm, height = 10cm]{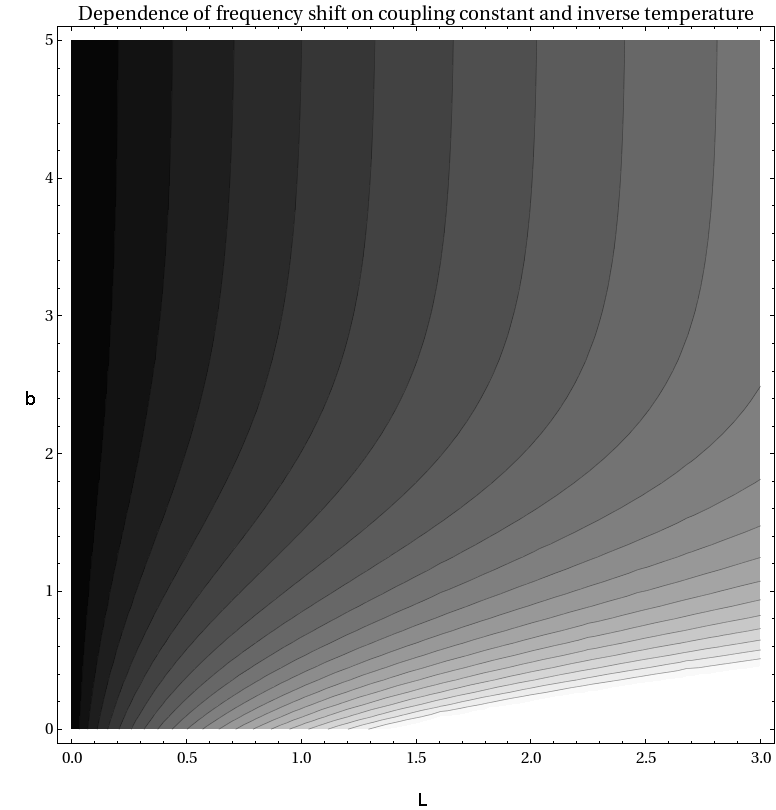}
\caption{Dependence of the $\Pi/\omega^{2}$ on dimensionless coupling constant 
$L = {\hbar} \lambda/(2 m \omega^{3})$ and dimensionless inverse temperature 
$b = \beta {\hbar} \omega$. The lighter regions correspond to larger values of
the $\Pi/\omega^{2}$}.
\end{center}
\end{figure}

Let us observe that the dependence of $\Pi$ on $\lambda$ is non-analytic even for 
zeroth temperature. Thus, the result is, in a sense, "non-perturbative".
 
\section{Statistical linearization of non-linear scalar field theory}
\label{Phi4}

We start with the following Euclidean action for the $\phi^{4}$ model:

\begin{equation}
S_{E} = \int_{0}^{\beta {\hbar}} d \tau \int d^3 x {\mathcal L}_{E},
\end{equation}

where

\begin{equation}
{\mathcal L}_{E} = 
\frac{1}{2 c^{2}} (\partial_{\tau} \phi)^{2} + \frac{1}{2}(\nabla \phi)^{2}  + 
\frac{1}{2 {\hbar}^{2}} m^2 c^2 \phi^{2} + \frac{g}{4!} \phi^{4}.
\end{equation}

The resulting Langevin equation for $\phi$ reads

\begin{equation}
\frac{\partial \phi}{\partial s} = \frac{1}{c^{2}} \frac{\partial^{2} \phi}{\partial \tau^{2}} + 
\nabla^{2} \phi - \frac{m^{2} c^{2}}{{\hbar}^{2}} \phi - \frac{g}{3!} \phi^{3} + \eta(s, t, {\bf r}). 
\end{equation}

We impose periodic boundary conditions in the Euclidean time, 
$\phi(s, 0, {\bf r}) = \phi(s, \beta {\hbar}, {\bf r})$. Also, as it is convenient to work with
discrete variables, we initially assume that $\phi$ is also periodic in each spatial variable,
that is, $\phi$ is defined in the box of the volume $L_{0}^{3}$ with periodic boundary
condition imposed in each spatial direction. At the end of calculations we shall take the limit
$L_{0} \rightarrow \infty$. Thus, we expand:

\begin{equation}
\phi(s, \tau, {\bf r}) = \frac{1}{\sqrt{\beta {\hbar} V}} \sum_{\lambda} \sum_{{\bf k}} 
\exp(-i \omega_{\lambda} \tau) \exp(i {\bf k} {\bf r}) \phi_{\lambda {\bf k}},
\end{equation}

where $\omega_{\lambda} = 2 \pi \lambda / \beta {\hbar}$, ${\bf k} = (2 \pi/L_{0}) (m , n, p)$,
and $\lambda$, $m$, $n$, $p$ are integers. 

Upon a similar Fourier decomposition of the stochastic forces we obtain:

\begin{equation}
\frac{d \phi_{\lambda {\bf k}}}{d s} = -\left( \frac{\omega_{\lambda}^{2}}{c^{2}} 
+ {\bf k}^{2} + \frac{m^{2} c^{2}}{{\hbar}^{2}} \right) \phi_{\lambda {\bf k}} - 
\frac{g}{3! \beta {\hbar} L_{0}^{3}} N_{\lambda {\bf k}} + \eta_{\lambda {\bf k}},
\end{equation}

where 

\begin{equation}
 N_{\lambda {\bf k}} = 
\sum_{\lambda_{1}, \lambda_{2}} \sum_{{\bf k}_{1}, {\bf k}_{2}}
\phi_{\lambda_{1}, {\bf k}_{1}} \phi_{\lambda_{2}, {\bf k}_{2}} 
\phi_{\lambda - \lambda_{1} - \lambda_{2}, {\bf k} - {\bf k}_{1} - \bf k_{2}}.
\end{equation}

The corresponding linear system takes the form:

\begin{equation}
\frac{d \phi_{\lambda {\bf k}}}{d s} = 
- A_{\lambda {\bf k}} \phi_{\lambda {\bf k}} + \eta_{\lambda {\bf k}}.
\end{equation}

Using the same prescription as before, i.e., minimization of the expectation value of the error, we find:

\begin{equation}
A_{\lambda {\bf k}} = 
\frac{\omega_{\lambda}^{2}}{c^{2}} 
+ {\bf k}^{2} + \frac{m^{2} c^{2}}{{\hbar}^{2}} + 
\frac{1}{2} \frac{g}{\beta {\hbar} L_{0}^{3}} 
\sum_{\lambda_{1} {\bf k}_{1}} \langle \phi_{\lambda_{1} {\bf k}_{1}} 
\phi_{-\lambda_{1}, -{\bf k}_{1}} \rangle
\end{equation}

On writing

$$
A_{\lambda {\bf k}} = \frac{\omega_{\lambda}^{2}}{c^{2}} 
+ {\bf k}^{2} + \frac{m^{2} c^{2}}{{\hbar}^{2}} + \Pi 
$$

and taking expectation values with respect to the Gaussian probability density
associated with the linear system, we obtain:

\begin{equation}
\Pi = \frac{g}{2 \beta L_{0}^{3}} \sum_{\mu} \sum_{{\bf k}}
\frac{1}{\omega_{\mu}^{2}/c^{2} + {\bf k}^{2} + m^{2} c^{2}/{\hbar}^{2} + \Pi},
\end{equation}

or by taking the limit $L_{0} \rightarrow \infty$,

\begin{equation}
\label{Phi4gap}
\Pi = \frac{g}{2 (2 \pi)^{3} \beta} \sum_{\mu} \int d^{3} k
\frac{1}{\omega_{\mu}^{2}/c^{2} + {\bf k}^{2} + m^{2} c^{2}/{\hbar}^{2} + \Pi}.
\end{equation}

This is a self-consistent, temperature-dependent equation for the self-energy
of the self-interacting non-linear scalar field. It agrees with that derived
in \cite{Altherr} (please see also \cite{LeBellac}).
Needless to say, the really serious analysis starts precisely at this point.
It must involve both the mass and coupling constant renormalization, and has
been performed, e.g., in \cite{DHLR,BIR}. It has been found that the 
separation of the zero-temperature and temperature-dependent terms as well as
determination of the thermal mass is surprisingly non-trivial, \cite{KR}.
In view of the fact that the matter has been carefully discussed in the above
papers, we shall not give any detailed analysis of our own. 
We would only like to investigate briefly the consequences of application
to Eq. (\ref{Phi4gap}) of the so-called Ramanujan technique of summation 
of divergent series.
But before doing this we would like to observe that, while there is nothing new
in Eq. (\ref{Phi4gap}), we have obtained it practically effortlessly
using the simplest, and actually trivial, version of statistical linearization
of stochastically quantized theory. No diagrammatic analyses or functional
differentiation or integration have been necessary.

Let us perform the frequency summation as well as the angular integration
in Eq. (\ref{Phi4gap}). The expression for $\Pi$ takes now the form:

\begin{eqnarray}
\Pi &=& \frac{g {\hbar} c}{8 \pi^{2}} \int_{0}^{\infty} \frac{k^{2} d k}
{\sqrt{k^{2} + m^{2} c^{2}/{\hbar}^{2} + \Pi}} \cdot
\nonumber \\
&& \cdot \left( 
1 + 2 n_{B} (c \beta {\hbar} \sqrt{k^{2} + m^{2} c^{2}/{\hbar}^{2} + \Pi})
\right).
\end{eqnarray}

The integral of the first term on the right-hand side is obviously quadratically
divergent. 

Let us have a closer look at the limit $\beta \rightarrow \infty$. The 
second term (which contains a convergent integral) on the right-hand side
vanishes in that limit. The self-consistent expression for the self-energy 
takes then the form:

\begin{eqnarray}
\Pi &=& \frac{g {\hbar} c}{8 \pi^{2}} \int_{0}^{\infty} \frac{k^{2} d k}{
\sqrt{k^{2} + (m^{2} c^{2}/{\hbar}^{2}) + \Pi}} = 
\nonumber \\
&& = \frac{g {\hbar} c p^{2}}{8 \pi^{2}}
\int_{0}^{\infty} \frac{\kappa^{2} d \kappa}{\sqrt{\kappa^{2} + 1}},
\end{eqnarray}

where $p^{2} = (m^{2} c^{2}/{\hbar}^{2}) + \Pi$ and we assume that $p^{2} > 0$.
Euler's substitution $\sqrt{\kappa^{2} + 1} = -\kappa + u$ yields:

\begin{eqnarray}
\label{diverge}
\Pi &=&  \frac{g {\hbar} c p^{2}}{32 \pi^{2}} \int_{1}^{\infty} 
\frac{(u^{2} - 1)^{2}}{u^{3}} = 
\nonumber \\
&& = \frac{g {\hbar} c p^{2}}{32 \pi^{2}}
\left( \int_{1}^{\infty} u du - 2 \int_{1}^{\infty} \frac{du}{u} + 
\frac{1}{2} \right).
\end{eqnarray}

In order to deal with the above divergent integrals let us invoke the Abel-Plana
formula \cite{CE}:

\begin{equation}
\int_{0}^{\infty} f(x) d x = \sum_{n=0}^{\infty} f(n) - \frac{1}{2} f(0)
- i \int_{0}^{\infty} \frac{f(i t) - f(-i t)}{e^{2 \pi t} - 1} dt.
\end{equation}

It makes sense if both the integrals and the series converge. However,
let us make use of the following definition. We say that an integral
$\int_{0}^{\infty} f(x) dx$ is summable in the Ramanujan sense if the 
sum of the series $\sum_{n=0}^{\infty} f(n)$ in the Ramanujan sense
\cite{Delabaere,Moreta} exists, and the integral 

$$
\int_{0}^{\infty} (f(i t) - f(-i t))/(e^{2 \pi t} - 1) dt
$$ 

converges in the ordinary sense. Then we write:

\begin{equation}
^{(R)}\int_{0}^{\infty} f(x) d x = ^{(R)}\sum_{n=0}^{\infty} f(n) - \frac{1}{2} f(0)
- i \int_{0}^{\infty} \frac{f(i t) - f(-i t)}{e^{2 \pi t} - 1} dt.
\end{equation}

with obvious meaning of the superscript $(R)$.

What is called here the ``Ramanujan summation" of divergent series consists in the following.
Let $\sum_{n \geq 1} a(n)$ be a formal (divergent) series, and let $R(x) = \sum_{n \geq 0} a(n+x)$.
Then, if $R(x)$ satisfies the difference equation $R(x) - R(x+1) = a(x)$, 
the value $R(1)$ gives the sum of the series $\sum_{n \geq 1} a(n)$. 
The sum is unique if we additionally require fulfillment of the condition
$\int_{1}^{2} R(x) = 0$ and $a(n)$ does not grow too fast with $n$. 
In \cite{Delabaere} an algorithm to compute $R(x)$ in terms of the Borel transform
of $a(n)$ as a well as a useful table of the Ramanujan sums of some divergent series is given.

With the above definitions in mind, we can investigate the question of summability of 
divergent integrals in Eq. (\ref{diverge}) in the Ramanujan sense.

We find immediately:

\begin{eqnarray}
^{(R)} \int_{1}^{\infty} u du &=& ^{(R)} \int_{0}^{\infty} u du - \frac{1}{2} = 
-\frac{1}{2} + ^{(R)} \sum_{n = 0}^{\infty} n + 
\nonumber \\
&& 
+ 2 \int_{0}^{\infty} \frac{t}{e^{2 \pi t} - 1} dt.
\end{eqnarray}

As the last integral is equal to $1/12$, we have:

\begin{equation}
^{(R)} \int_{1}^{\infty} u d u = -\frac{5}{12} + ^{(R)} \sum_{n = 0}^{\infty} n.
\end{equation}

But, according to \cite{Delabaere},

$$
^{(R)} \sum_{n = 1}^{\infty} n^{k} = \frac{1 - B_{k+1}}{k + 1},
$$

where $B_{k}$ are the Bernoulli numbers. Hence $^{(R)} \sum_{n = 1}^{\infty} n = 5/12$, and 

$$
^{(R)} \int_{1}^{\infty} u du = 0.
$$

What is more,

\begin{equation}
^{(R)} \int_{1}^{\infty} \frac{du}{u} = 
^{(R)} \sum_{n = 0}^{\infty} \frac{1}{n + 1} - \frac{1}{2} - 
2 \int_{0}^{\infty} \frac{t}{(t^{2} + 1) (e^{2 \pi t} - 1}) dt.  
\end{equation}

Using now the fact that in the Ramanujan sense the sum of the harmonic series is the Euler-Mascheroni
constant $\gamma$ while

$$
-2 \int_{0}^{\infty} \frac{t}{(t^{2} + z^2) (e^{2 \pi t} - 1)} dt = \psi(z) + \frac{1}{2 z} - log(z),
$$

where $\psi(z)$ is the digamma function \cite{Lebedev} (please see the formula 6.3.21 of \cite{AS}), 
we find that, quite trivially,

\begin{equation}
^{(R)} \int_{1}^{\infty} \frac{du}{u} = 0,
\end{equation}

because $\psi(1) = -\gamma$.

As a result of the Ramanujan summation, we get in the zeroth-temperature limit:

\begin{equation}
\Pi = \frac{g {\hbar} c}{64 \pi^{2}} \left( \frac{m^{2} c^{2}}{{\hbar}^{2}} + \Pi \right),
\end{equation}

or 

\begin{equation}
\Pi = \frac{g_{ren} {\hbar} c}{64 \pi^{2}} \frac{m^{2} c^{2}}{{\hbar}^{2}}
\end{equation}

with $g_{ren} = g/(1 - (g {\hbar} c/64 \pi^{2}))$.

Thus the ``renormalization" of the coupling constant appears to be finite (provided that
$g {\hbar c} \neq 64 \pi^{2}$), and the same is true about the correction to the mass.
This is possible only because we have performed a summation of two divergent
integrals. In the present case, the Ramanujan approach resulted in their trivial elimination.

We would like to repeat that the formal manipulation with divergent series in the 
spirit of Ramanujan cannot be thought of as a substitution of the detailed and careful 
renormalization procedure. Nonetheless, the former is perhaps of non-vanishing  
interest and may be worth of some further study.

\section{Bose-Hubbard model}

Let us now consider the one-dimensional Bose-Hubbard model. 
It is of some interest because of the potential application
of cold Bose gases on lattices as an ``analog computer" to simulate properties of gauge fields.

The Bose-Hubbard model describes the (thermo)dynamics of Bose particles on a lattice
under the assumption that the particles can jump only between neighboring sites,
and only the on-site interactions are present. Mathematically, it is defined by the following 
(real-time) action:

\begin{eqnarray}
S &=& \int_{t_{1}}^{t_{2}} d t \sum_{m} \left[ \frac{i {\hbar}}{2} \left( \alpha_{m}^{\star} 
\frac{d \alpha_{m}}{d t} - \frac{d \alpha_{m}^{\star}}{d t} \alpha_{m} \right)
\right.
\nonumber \\
&& \left. + J \left( \alpha_{m}^{\star} \alpha_{m+1} + \alpha_{m+1}^{\star} \alpha_{m} \right) \right.
\nonumber \\
&& \left. + \mu_{c} \alpha_{m}^{\star} \alpha_{m} - \frac{V}{2}
\alpha_{m}^{\star 2} \alpha_{m}^{2} \right],
\end{eqnarray}

where $J$ is the so-called hopping parameter, $V$ is the interaction strength, and $\mu_{C}$ is the 
chemical potential. The index $m$ enumerates the sites on one-dimensional lattice, and we assume
that size of the lattice is $N$.

In the present case, it more convenient to work in real time, and perform the analytic continuation
to imaginary time only at the end of calculations. 
Let us notice that the well-known difficulties in the simulations of the Langevin equations
in the real time do not concern us here because the whole procedure is analytic (except of the last
step).

Using the prescription by \cite{HR}
and \cite{Gozzi}, we write

\begin{equation}
\frac{d \alpha_{m}}{d s} = i \frac{\delta S}{\delta \alpha_{m}^{\star}} + \eta_{m}(s, t), \;\;\;\;\; 
\frac{d \alpha_{m}^{\star}}{d s} = i \frac{\delta S}{\delta \alpha_{m}} + \eta_{m}^{\star}(s, t).
\end{equation}

The complex amplitudes are, however, to be considered as independent quantities, so that we effectively
complexify the theory.

Taking the functional derivatives gives us the system:

\begin{equation}
\frac{\partial \alpha_{l}}{\partial s} = -{\hbar} \frac{\partial \alpha_{l}}{\partial t} + 
i J (\alpha_{l+1} + \alpha_{l-1}) + i \mu_{c} \alpha_{l} - i V \alpha_{l}^{\star} \alpha_{l}^{2}
- \epsilon \alpha_{l} + \eta_{l}, 
\end{equation}

\begin{equation}
\frac{\partial \alpha_{l}^{\star}}{\partial s} = {\hbar} \frac{\partial \alpha_{l}^{\star}}{\partial t} + 
i J (\alpha_{l+1}^{\star} + \alpha_{l-1}^{\star}) + i \mu_{c} \alpha_{l}^{\star} 
- i V \alpha_{l}^{\star 2} \alpha_{l} - \epsilon \alpha_{l}^{\star} + \eta_{l}^{\star}. 
\end{equation}

In the above equations $\epsilon$ denotes an infinitesimal positive constant introduced to guarantee
the existence of stationary solutions. The stochastic forces satisfy the relation:

\begin{equation}
\langle \eta_{l}(s, t) \eta_{l^{\prime}}^{\star}(s^{\prime}, t^{\prime} \rangle = 2 {\hbar} \delta_{l l^{\prime}}
\delta(s - s^{\prime}) \delta(t - t^{\prime}).
\end{equation}

We impose upon $\alpha_{l}$ and $\alpha_{l}^{\star}$ periodic boundary conditions in time $t$ 
with the period $T_{0}$. 
We shall also assume that the whole lattice is periodic, that is $\alpha_{l} = \alpha_{N + l}$
for some integer $N$. This allows us to diagonalize the linear parts of the above equations
by expanding:

$$
\alpha_{l} = \frac{1}{\sqrt{N T_{0}}} \sum_{k = 0}^{N-1} \sum_{\mu = -\infty}^{\infty}
\exp(2 \pi i k l/N) \exp(-2 \pi i \mu t/T_{0}) f_{k \mu}(s),
$$

and analogously for $\alpha_{l}^{\star}$, $\eta_{l}$, $\eta_{l}^{\star}$.

The complex amplitudes $f_{k \mu}$ satisfy the relations:

\begin{eqnarray}
\frac{\partial f_{k \mu}}{\partial s} &=& \left[ i {\hbar} \omega_{\mu} + 2 i J \cos(P_{k}) + i \mu_{c}
- \epsilon \right] f_{k \mu} 
\nonumber \\
&& - i \frac{V}{T_{0} N} \sum_{p, q} \sum_{\nu, \rho}
f_{p \nu}^{\star} f_{q \rho} f_{k + p - q, \mu + \nu - \rho} + \xi_{k \mu},
\end{eqnarray}

\begin{eqnarray}
\frac{\partial f_{k \mu}^{\star}}{\partial s} &=& \left[ i {\hbar} \omega_{\mu} + 2 i J \cos(P_{k}) + i \mu_{c}
- \epsilon \right] f_{k \mu}^{\star}
\nonumber \\
&& - i \frac{V}{T_{0} N} \sum_{p, q} \sum_{\nu, \rho}
f_{p \nu}^{\star} f_{q \rho} f_{k - p + q, \mu - \nu + \rho} + \xi_{k \mu}^{\star},
\end{eqnarray}

where $\langle \xi_{k \mu}(s) \xi_{k^{\prime} \mu^{\prime}}^{\star}(s^{\prime}) \rangle = 
2 {\hbar} \delta_{k k^{\prime}} \delta_{\mu \mu^{\prime}} \delta(s - s^{\prime})$,
and $P_{k} = 2 \pi k / N$.

The equivalent linear systems read:

\begin{equation}
\frac{\partial f_{k \mu}}{\partial s} = - i A_{k \mu} f_{k \mu} + \xi_{k \mu}(s),
\end{equation}

\begin{equation}
\frac{\partial f_{k \mu}^{\star}}{\partial s} = -i A_{k \mu}^{\star} f_{k \mu}^{\star} 
+ \xi_{k \mu}^{\star}(s),
\end{equation}

where $A_{k \mu}$ and $A_{k \mu}^{\star}$ are not complex-conjugated quantities.

Minimization of the error leads now to the following self-consistent equation:

\begin{equation}
A_{k \mu} = - {\hbar} \omega_{\mu} - 2 J \cos(P_{k}) - \mu_{c} - i \epsilon
+  \sum_{p, \nu} \frac{2 {\hbar}}{i A_{p \nu}},
\end{equation}

or,

\begin{equation}
A_{k \mu} = - {\hbar} \omega_{\mu} - 2 J \cos(P_{k}) - \mu_{c} - i \epsilon
+ \Pi,
\end{equation}

where

\begin{equation}
\Pi = \frac{2 {\hbar} V}{i N T_{0}} \sum_{p, \nu} \frac{1}{\Pi - {\hbar} \omega_{\nu}
- 2 J \cos(P_{p}) - \mu_{c} - i \epsilon}.
\end{equation}

We now make the rotation in the complex plane to obtain the results in imaginary time and 
write simply $T_{0} = -i \beta {\hbar}$. Summation over $\nu$ can be easily performed to give
the self-energy as a sum over finite number of discrete momenta:

\begin{equation}
\label{BHSE}
\Pi = \frac{V}{N} \sum_{p=0}^{N-1} \coth \left((\beta/2) (\Pi - 2 J \cos(P_{p}) - \mu_{c}) \right).
\end{equation}

Let us notice that the zeroth-temperature limit is not obvious because we do not know the sign
of the expression inside the hyperbolic cotangent function.
The solutions to the above self-consistent equation for $\Pi$ provide us, upon the 
analytic continuation from Matsubara's to the retarded Green function, with the 
temperature-dependent approximation to the self-energy of a particle in the Bose-Hubbard model. 
While the analysis of Eq. (\ref{BHSE}) is beyond the scope of the present work,
we provide Fig. 2 which illustrates the general structure of solutions.
In this figure we have plotted curves representing the function

$$
y = f(x) = \frac{1}{N} \sum_{p = 0}^{N-1} \coth \left((v/2) (x - 2 j \cos(P_{p}) - u) \right)
$$

(i.e., the right-hand side of (\ref{BHSE})), where $v = \beta V$, $j = J/V$, $u = \mu_{c}/V$ as
dependent on $x = \Pi/V$, with superimposed straight line $y = x$. The points at which that straight line 
crosses the curves represent solutions of (\ref{BHSE}). The parameters chosen to plot Fig. 2
have been: $j = 0.1$, $u = 0.5$, $v = 1$ and $20$.

\begin{figure}
\begin{center}
\includegraphics[width = 6.5cm, height = 7cm]{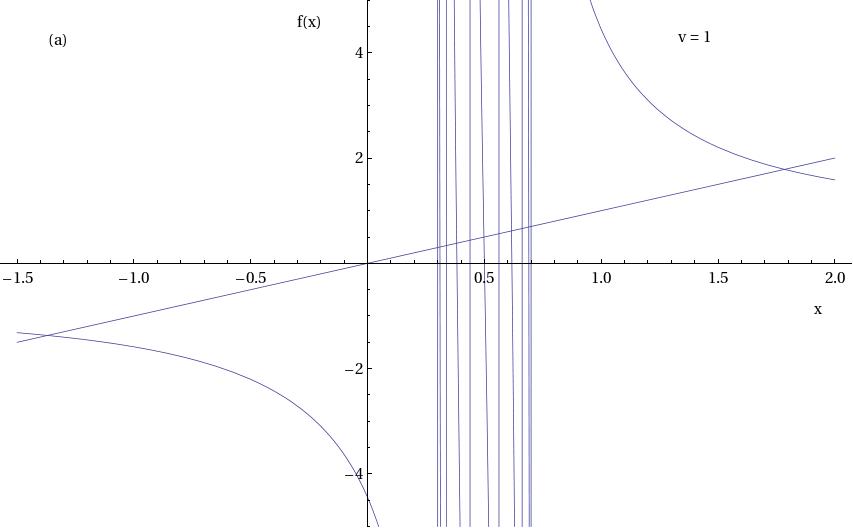}
\includegraphics[width = 6.5cm, height = 7cm]{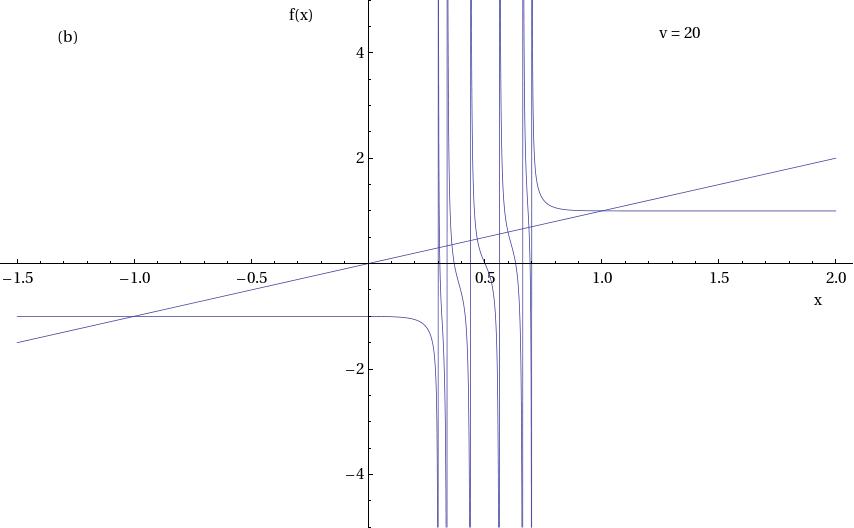}
\caption{Graphical solution to the self-consistent equation for the self-energy
in the Bose-Hubbard model for $J/V = 0.1$, $\mu/V = 0.5$, N = 10. 
The solutions (i.e. the values of $\Pi/V$ which satisfy (\ref{BHSE})) are given 
by the points where the curves are crossed by the straight line;
(a) $\beta V = 1$, (b) $\beta V = 20$}.
\end{center}
\end{figure}

It is to be noticed that one of the parameters we used to plot Fig. 2, namely
$j = J/V$ has been taken to be $0.1$. Such a value is quite realistic.
However, it obviously means that we are in the strong coupling regime, for which
the statistical linearization may lead to erroneous results. Thus, the above 
figures are to be understood merely as an illustration of possible solutions
of (\ref{BHSE}). There is, unfortunately, no general theory which could determine
whether or not the statistical linearization provides valid
approximations in the strong coupling regime.

As is clearly seen from Eq. (\ref{BHSE}) and from the figures, the gap equation
possesses multiple solutions. A similar situation occcurs in both QCD \cite{WQLCRS} 
and in the superconductivity theory \cite{SSL}. We plan to discuss the important
and interesting question of stability of various solutions elsewhere.

\section{Concluding remarks}
In this paper we have applied the statistical linearization technique to the stochastically
quantized field equations in Euclidean setting. Three examples have been given: 
the quantum anharmonic oscillator,
the real scalar $\phi^{4}$ theory, and the Bose-Hubbard model. In all three cases a self-consistent
non-perturbative expression for the temperature-dependent self-energy has been provided. 
Only the simplest, Gaussian, version of the linearization technique have been employed.
As a matter of fact, the statistical linearization methods are very rich. At least two ways
can be used in order to improve the acuracy of results. Firstly, one can apply the so-called
higher-order linearization \cite{Iyengar,RS,Socha}. Within that approach, one replaces non-linear
terms in the field equations by a new dependent variable, which satisfies its own (complicated)
differential equation; the latter is then linearized by error minimization. That method
is somewhat analogous to, but not identical with, the breaking of the Schwinger-Dyson 
(or, analogously, BBGKY) hierarchy of the integro-differential equations for the correlation
functions.   
Secondly, one can also employ the so-called ``equivalent non-linearization'' \cite{Caughey2,RS,Socha} 
approach in which the original non-linear equations are replaced by simpler (though still non-linear)
equations with known statistical properties of solutions.
Work is in progress on application of both those techniques in the physics of cold gases,
quantum electrodynamics and quantum optics.

\end{document}